\documentclass[aps,nofootinbib,twocolumn,notitlepage,superscriptaddress, amsmath,amssymb,amsfonts,floatfix]{revtex4}

\usepackage{amsmath}
\usepackage{amssymb}
\usepackage{amsfonts}
\usepackage[dvips]{graphicx}
\usepackage{epsfig}
\usepackage{tabularx}
\usepackage{color}

\newcommand{\Omh}{\Omega_\mathrm{H}}
\newcommand{\Omz}{\Omega_\mathrm{0}}
\newcommand{\Omegai}{\Omega_\mathrm{I}}
\newcommand{\Omegar}{\Omega_\mathrm{R}}
\newcommand{\rz}{r_\mathrm{0}}
\newcommand{\rh}{r_\mathrm{H}}
\newcommand{\nz}{n_\mathrm{0}}
\newcommand{\nd}{n_\mathrm{2}}

\begin{document}

\title{Acoustic black-hole bombs and scalar clouds in a photon-fluid model}
\author{Marzena Ciszak}
	\affiliation{CNR-Istituto Nazionale di Ottica, Via Sansone 1, I-50019 Sesto Fiorentino (FI), Italy.}
	\author{Francesco Marino}
	\affiliation{CNR-Istituto Nazionale di Ottica, Via Sansone 1, I-50019 Sesto Fiorentino (FI), Italy.}
	\affiliation{INFN, Sezione di Firenze, Via Sansone 1, I-50019 Sesto Fiorentino (FI), Italy}

\date{\today}

\begin{abstract}

Massive bosonic fields in the background of a Kerr black hole can either trigger superradiant instabilities (black-hole bombs) or form equilibrium configurations corresponding to pure bound states, known as stationary scalar clouds. Here, similar phenomena are shown to emerge in the fluctuation dynamics of a rotating photon-fluid model. In the presence of suitable vortex flows, the density fluctuations are governed by the \emph{massive} Klein-Gordon equation on a (2+1) curved spacetime, possessing an ergoregion and an event horizon. We report on superradiant instabilities originating from quasi-bound phonon states trapped by the vortex background and, remarkably, on the existence of stationary modes in synchronous rotation with the horizon. These represent the acoustic counterpart of astrophysical scalar clouds. Our system offers a promising platform for analogue gravity experiments on superradiant instabilities of massive bosons and black-hole-field equilibrium configurations.

\end{abstract}

\maketitle

\section{Introduction}

Superradiant scattering is an effect whereby waves reflected at the interface of a moving medium are amplified, extracting energy and momentum in the process. For an axially-symmetric rotating system the amplification occurs whenever
\begin{equation}
\Omega < m \Omega_\textrm{C}
\label{eq1}
\end{equation}
where $\Omega$ is the angular frequency of the incident wave, $m$ its azimuthal wavenumber with respect to the rotation axis and $\Omega_\textrm{C}$ the angular velocity of the rotating object. This phenomenon was originally introduced by Zel'dovich in 1971 \cite{zeldovich} who investigated the scattering of electromagnatic waves from a rotating conducting cylinder. Similarly, a bosonic field impinging upon a Kerr Black Hole (BH) is expected to be amplified if the superradiance condition (\ref{eq1}) holds, in which case the critical frequency $\Omega_\textrm{C}=\Omh$ is the angular velocity at the event horizon \cite{misner}.
According to quantum mechanics, rotating absorbing bodies -including BHs- would support the spontaneous emission of superradiant particles, being slowed down by the process \cite{zeldovich,staro1,staro2}. The possibility to extract energy from a spinning BH was quantified by Penrose some years before \cite{penrose}, and it is related to particles entering the ergosphere with negative energies\footnote{in a reference frame compatible with that of a Minkowskian observer far from the black hole.}. This work and the above studies of quantum pair production, can be hystorically considered as precursors to Hawking's result on BH evaporation \cite{hawking}.

In the presence of a confining mechanism, superradiant effects have profound consequences on the stability properties of spinning BHs. An initially small perturbation with frequency in the superradiant regime will grow exponentially in time, if it is prevented from radiating its energy to infinity. This is the so-called \emph{black-hole bomb} by Press and Teukolsky \cite{bhbomb}. In their proposal the confinement is provided by a perfectly reflecting mirror surrounding the rotating BH: a field thus undergoes repeated superradiant scattering between the horizon and the mirror, resulting in the exponential growth of the field amplitude and instability of the system. This is expected e.g. in Kerr--Anti--de Sitter (AdS) spacetimes where the conformal boundary can act as an effective mirror \cite{Kerrads}.

Remarkably, similar phenomena occurs when a Kerr BH is simply coupled to a massive scalar field \cite{damour,zouros,detw,furu,cardoso1,dolan1}. The gravitational interaction between the BH and the field gives rise to a binding potential that prevents low-frequency modes in the regime 
\begin{equation}
\Omega < \Omz
\label{eq2}
\end{equation}
from escaping to spatial infinity, where $\Omz$ is the rest frequency of the massive field.
The potential gives rise to the formation of a discrete set of quasi-bound (non-stationary) states in the vicinity of the horizon characterized by complex eigenfrequencies $\Omega$. Modes further satisfying the condition (\ref{eq1}), $\Omegar=\mathrm{Re} (\Omega) < m \Omh$, are superradiantly amplified thus triggering the BH bomb mechanism. On the other hand, for $\Omegar > m \Omh$ the modes are decaying in time, which indicates that the field is infalling into the horizon and the system is linearly stable. 

The boundary between the stable and unstable regime $\Omegar=m \Omh$ is marked by field configurations which are in equilibrium with the BH, i.e. $\Omegai=\mathrm{Im} (\Omega)$=0. These stationary (infinitely long-lived) modes form {\it pure} bound states in synchronous rotation with the horizon, known as scalar clouds \cite{hod1,hod2}.

The study of such configurations is of paramount interest in BH theory: their existence at the linear level is intimately linked to {\it hairy} black hole solutions of the (nonlinear) Einstein-Klein-Gordon system \cite{herdeiro1} and they are considered as viable dark-matter candidates \cite{hod1}. Upper bounds to the clouds mass in Kerr spacetime have been derived \cite{hod3}, and recently, stationary charged clouds have been shown to exist around Kerr-Newman BHs \cite{benone,hod4,Huang}. On the other hand, superradiant instabilities have potential implications in cosmology and high-energy physics \cite{reviewSR}, for instance in the discovery of primordial BHs through their coupling with massive bosons or, conversely, of yet-unknown ultralight particles (e.g. axions) interacting with supermassive BHs \cite{urbano}.

Here, we show that similar phenomena can be observed in the excitation dynamics of a rotating photon-fluid in the presence of both local and nonlocal interactions.

Photon-fluids are nonlinear optical systems which can be described in terms of the hydrodynamic equations of an interacting Bose gas \cite{rica1992,rica1993}. Similarly to classical \cite{unruh,visser1} and quantum fluids \cite{volovik,garay2001} (for a review see \cite{rev}), long-wavelenght excitations (phonons) in an inhomogeneous flow propagate as a massless scalar field on a curved spacetime endowed with a Lorentzian metric (acoustic metric) \cite{marino2008}. These systems are an ideal platform for analogue gravity investigations \cite{barad,fouxon,elazar,marino2016}. In particular, superradiant scattering in rotating photon-fluids has been theoretically demonstrated \cite{marino2009,ornigotti,prainSR,braidottiPRL}, and recently a rotating BH geometry has been experimentally realized \cite{vocke2018}.

Interestingly, photon-fluids with both local and nonlocal optical nonlinearities support massive elementary excitations which, at lower momenta, correspond to massive phonons with a relativistic energy-momentum relation \cite{marino2019}. 
In this system, we investigate the phonon dynamics over a draining vortex flow, as the acoustic analogue of a massive scalar field on a rotating BH. We report on superradiant instabilities, originating from the existence of quasi-bound phonon states trapped in the vicinity of the horizon and, remarkably, scalar cloud solutions, equilibrium configurations between the massive phononic field and the vortex background.
Similar states in analogue models has been previously discussed by Benone {\it et al.} \cite{benone-ac}, who considered a rotating acoustic BH enclosed in a cylindrical cavity. Unlike "real" clouds which occur for massive fields, phonons in standard fluids have no mass and a cavity is required to confine them in the vicinity of the horizon. In our model instead, phononic excitations propagate as a \emph{massive} Klein-Gordon field, and are naturally trapped by a binding potential in close analogy to astrophysical scenarios.

The paper is organized as follows. In Sec. II, we introduce the photon-fluid model making the connection between hydrodynamic and optical quantities. In Sec. III we review how the massive Klein-Gordon equation governing the propagation of density fluctuations is derived in the long-wavelength limit. In Sec. IV we decompose the field to derive the radial Teukolsky equation in the vortex metric, which will be the starting point of our analysis. We then show in Sec. V that perturbations with $\Omega< m \Omh$ are superradiantly amplified, producing either superradiant scattering when $\Omega>\Omz$ or dynamical instabilities when $\Omega<\Omz$. In this second case, illustrated in Sec. VI, quasi-bound states are shown to exist for a discrete set of complex frequencies. Finally, in Sec. VII we focus on scalar clouds solutions and calculate numerically the existence lines in the BH-field parameters space ($\Omh$,$\Omz$). Conclusions and future perspectives are presented in Sec. VIII.

\section{Photon-fluid and collective excitations}

The propagation of a monochromatic optical beam oscillating at angular frequency $\omega$ in a 2D nonlinear medium can be described within the paraxial approximation in terms of the Nonlinear Schr\"{o}dinger Equation \cite{boyd} 
\begin{equation}
\partial_z E = \frac{i}{2 k} \nabla^{2} E - i \frac{k}{\nz} E \Delta n \;
\label{eq3}
\end{equation}
where $E$ is the slowly varying envelope of the electromagnetic field, $z$ is the propagation coordinate, $k=2 \pi \nz /\lambda$ is the wavenumber, $\lambda$ the vacuum wavelength and $\nz$ is the linear refractive index. The laplacian term $\nabla^{2} E$ defined with respect to the transverse coordinates ${\bf r}=(x,y)$ accounts for diffraction and $\Delta n$ is the nonlinear optical response of the medium. For a purely local (Kerr-defocusing) nonlinearity \cite{note}, $\Delta n = \nd \vert E \vert^{2}$ with $\nd>0$, Eq. (\ref{eq3}) is formally identical to the 2D Gross-Pitaevskii equation for a dilute boson gas with repulsive contact interactions \cite{bec}. The dynamics takes place in the transverse plane $(x,y)$ of the laser beam so that the propagation coordinate $z$ plays the role of an effective time variable $t=(\nz/c)z$, where $c$ is speed of light in vacuum.

Here we consider an optical medium with both local Kerr and nonlocal thermo-optical nonlinearities $\Delta n = \nd \vert E \vert^{2} + \mathrm{n_{th}}$. The refractive index change $\mathrm{n_{th}}$ is thus coupled to the optical intensity through the stationary heat equation \cite{Swartz,carmon}
\begin{equation}
- \nabla^{2} \mathrm{n_{th}} = \frac{\alpha \beta}{\kappa} \vert E \vert^{2}
\label{eq4}
\end{equation}
where $\kappa$ is the thermal conductivity of the material, $\alpha$ its linear absorption coefficient and $\beta = \partial \mathrm{n_{th}} / \partial T >0$ is the change in the refractive index with respect to the temperature\footnote{Eq. (\ref{eq4}) holds in the limit of an infinite medium in the two transverse dimensions, and implies an infinite-range for the nonlocal thermo-optical interactions. More realistic models take into account the effects of spatial boundaries by including loss terms in the stationary heat equation, which make the length-scale of the thermo-optical nonlinearity finite \cite{vocke2016,conti1,conti2}.}. Nevetheless, a regime infinite-range nonlocality can be reasonably reproduced by means of suitable background optical beams \cite{danieleSN}. Similar models arise in semiconductor optical materials \cite{torner}, nematic liquid crystals \cite{warenghem} and appear also in Bose-Einstein Condensates (BECs) with simultaneous local and long-range (e.g. dipolar) interactions \cite{dipolar}.

The corresponding hydrodynamic formulation of (\ref{eq3}-\ref{eq4}) is obtained by means of the Madelung transform $E = \rho^{1/2} e^{i \phi}$,
\begin{eqnarray}
\partial_t \rho + {\bf \nabla}\cdot (\rho {\bf v}) = 0 \label{eq5a}\;  \\
\partial_t \psi + \frac{1}{2} v^2 = -\frac{c^2}{\nz^3}n_2 \rho - \frac{c^2}{\nz^3} \mathrm{n_{th}} + \frac{c^2}{2 k^2 \nz^2}\frac{\nabla^2 \rho^{1/2}}{\rho^{1/2}} \label{eq5b} \, 
\end{eqnarray}
where the optical intensity $\rho$ corresponds to the fluid density and ${\bf v} = \frac{c}{k \nz}\nabla \phi \equiv \nabla \psi$ is the flow velocity. On the right-hand side of (\ref{eq5b}), the first term provides the local repulsive interactions related to the positive bulk pressure $P =\frac{c^2 n_2}{2 n_0^3} \rho^2$, $\mathrm{n_{th}}$ gives rise to a nonlocal interaction potential, while the last term, directly related to diffraction, is the analogue of the Bohm quantum potential.

\subsection{Bogoliubov-De Gennes equations and dispersion relation}

The evolution of the first-order complex fluctuations $\varepsilon({\bf r},t)$ of the optical field is obtained by linearizing Eq. (\ref{eq1}) around a background solution, $E = E_0(1 + \varepsilon + ...)$ with $E_0=\rho_0^{1/2} e^{i \phi_0}$. We obtain
\begin{eqnarray}
(\partial_T -i \frac{c}{2 k n_0} \partial_S) \varepsilon = - i \frac{\omega}{n_0} (n_2 \rho_0(\varepsilon + \varepsilon^*) + n_{th}) \label{eq6a}\;  \\
(\partial_T +i \frac{c}{2 k n_0} \partial_S) \varepsilon^* = i \frac{\omega}{n_0} (n_2 \rho_0(\varepsilon + \varepsilon^*) + n_{th}) \label{eq6b} \;  \\
-\nabla^2 n_{th} = \frac{\alpha \beta}{\kappa} \rho_0(\varepsilon + \varepsilon^*) \label{eq6c} \;
\end{eqnarray}
in which we introduce the comoving derivative $\partial_T = \partial_t + \bf{v_0} \cdot \nabla$ with ${\bf v_0}=\frac{c}{k n_0} \nabla \phi_0$, and the spatial differential operator $\partial_S = \frac{1}{\rho_0} \nabla \cdot (\rho_0 \nabla ~ ~)$. 
Eq. (\ref{eq6c}) defines the refractive-index fluctuations due to thermo-optic effect, $n_{th}$. For $n_{th}=0$, Eqs. (\ref{eq6a}-\ref{eq6b}) takes the form of the standard Bogoliubov-De Gennes equations governing the dynamics of elementary excitations in purely local quantum fluids \cite{bec}.

In the spatially homogeneous case where both the background density $\rho_0$ and velocity $\bf{v_0}$ do not depend on the transverse coordinates, the plane-wave solutions of Eqs. (\ref{eq6a})-(\ref{eq6c}) satisfy the dispersion relation
\begin{equation}
\Omega^2 = \Omz^2 + c_s^2 K^2\left(1 + \frac{\xi^2}{4\pi^2}K^2\right)
\label{disp1} 
\end{equation}
where $\bf{K}$ is the transverse wavevector of the mode ($K$ is the wavenumber) $\Omega = \Omega' - {\bf K}\cdot{\bf v_0}$ its angular frequency in the locally-comoving background frame and $\Omz=c \sqrt{\frac{\alpha \beta}{\kappa n_0^3} \rho_0}$. In analogy to purely local BECs and photon-fluids \cite{marino2008}, we define the sound speed as $c_s^2 \equiv \frac{d P(\rho_0)}{d \rho} = \frac{c^2 n_2}{n_0^3} \rho_0$ and the healing length, $\xi=\lambda/2\sqrt{n_0 n_2 \rho_0}$ as the characteristic length separating the linear (phononic) and quadratic (single-particle) regime of the dispersion relation (\ref{disp1}). The length $\xi$ thus determines the critical wavenumber $K_c=2\pi/\xi$ associated to the breakdown of Lorentz invariance. Low energy modes with $K\ll K_c$ (phonons) obey the relativistic dispersion relation for a massive particle, where $\hbar \Omz = m_\textrm{ph} c_s^2$ identifies the rest energy of the collective excitations, $m_\textrm{ph}$ the rest mass and $c_s$ plays the role of the light speed. At higher momenta $K \gg K_c$, the terms related to the Bohm quantum potential become dominant and the excitations propagate with a group velocity increasing with $K$. Similar high-energy (Lorentz-violating) corrections appear in several phenomenological approaches to quantum gravity, where $\hbar K_c$ is typically associated to the Planck momentum \cite{slv}.

\section{Phonons as a Massive Klein-Gordon field}

The formal equivalence between phonons propagating on top of an inhomogeneous photon-fluid and the evolution of scalar fields in curved spacetime can be established starting from the Bogoliubov-de Gennes equations (\ref{eq6a})-(\ref{eq6b}) in the limit $K \ll K_c$ \cite{marino2019}.
To this end, we apply the operator $(\partial_T +i \frac{c}{2 k n_0} \partial_S) (\frac{1}{\rho_0} ~ ~)$ to Eq. (\ref{eq6a}) and we obtain
\begin{eqnarray}
(\partial_T +i \frac{c}{2 k n_0} \partial_S)\frac{1}{\rho_0}(\partial_T -i \frac{c}{2 k n_0} \partial_S) \varepsilon = \frac{c_s^2}{\rho_0} \partial_S \varepsilon  \nonumber\\ 
- i (\frac{\omega}{n_0}\partial_T +i \frac{c^2}{2 n_0^3} \partial_S)\frac{1}{\rho_0} n_{th}
\label{eq7}
\end{eqnarray}
The phonon dynamics is obtained by ignoring higher-order spatial derivatives in Eq. (\ref{eq7}) (\emph{long-wavelength limit}), which indeed are responsible for the Lorentz-breaking, $K^4$-terms in the dispersion relation \cite{prainSR,marino2019}. We will turn back on the implications of this approximation at the end of Sect. VII.
In this limit and using the fact that the background density $\rho_0$ satisfies the continuity equation (\ref{eq5a}) with ${\bf v}$=${\bf v_0}$, Eq. (\ref{eq7}) can be rewritten as 
\begin{eqnarray}
\Box \varepsilon - \frac{1}{2} \Omz^2 (\varepsilon + \varepsilon^*)\,= \, i \frac{\omega}{n_0} (\partial_T+\nabla\cdot {\bf v_0}) n_{th} \nonumber\\ 
+ \frac{c^2}{2 n_0^3}\nabla \cdot (n_{th} \nabla \ln \rho_0) 
\label{eq8}
\end{eqnarray}
where 
\begin{equation}
\Box \equiv - (\partial_T+\nabla\cdot {\bf v_0})\partial_T+\nabla \cdot(c_s^2\nabla \,\,\,)
\label{eq8b}
\end{equation}
is the d'Alambertian operator associated with the acoustic metric
\begin{equation}
g_{\mu\nu} =
\left(\frac{\rho_0}{c_s} \right)^2 \left(\begin{array}{cc}
  -(c_s^2 - v_0^2)  &  -{\bf v_0^T} \\
  -{\bf v_0}  &  {\bf I} \\
\end{array}
\right)
\label{metric-a}
\end{equation}
and ${\bf I}$ stands for the two-dimensional identity matrix.

The complex fluctuations $\varepsilon$ can be easily linked to the real density and phase perturbations through the relations $\rho_1=\rho_0(\varepsilon+\varepsilon^*)$ and $\phi_1=(i/2)(\varepsilon^*-\varepsilon)$. By means of these expressions and using $\psi_1= (c/k n_0) \phi_1$, Eq. (\ref{eq8}) splits into the following system of coupled wave equations 
\begin{eqnarray}
\Box \left(\frac{\rho_1}{\rho_0}\right) - \Omz^2 \frac{\rho_1}{\rho_0} = \frac{c^2}{n_0^3} \nabla \cdot (n_{th} \nabla \ln \rho_0) \label{eq10a}\;  \\
\Box \psi_1 = \frac{c^2}{n_0^3} (\partial_T + \nabla \cdot {\bf v_0}) n_{th} \label{eq10b}
\end{eqnarray}
for the fluctuations in the velocity-potential $\psi_1$ and the relative density $\rho_1/\rho_0$. 
Notably, the above equations decouple for a nearly-homogeneous background density $\rho_0 \approx const$. On the basis of Eqs. (\ref{eq5a})-(\ref{eq5b}), a nearly-constant density would also imply a nearly-homogeneous flow. Nevertheless, there are specific situations in which the simplifying assumption to consider a nearly-constant density while letting the flow vary provides an useful and realistic model of the system. This is indeed the case of vortex flows in which the horizon and the ergosphere are located away from the vortex core in regions of nearly constant density \cite{marino2008,vocke2018}. 
In these conditions Eq. (\ref{eq10b}) takes the form of the massive Klein-Gordon equation in curved spacetime
\begin{equation}
\Box \rho_1 - \Omz^2 \rho_1 = 0 \,.
\label{eq11}\;  \\
\end{equation}

\section{Teukolsky equation in the vortex metric}

Superradiant instabilities and scalar clouds require (at least) a rotating Kerr geometry that in acoustic models can be simulated by draining vortex flows. In photon-fluids, a vortex arises from the self-trapping of a phase singularity embedded in a broad optical beam due to the counterbalanced effects of self-defocusing and diffraction \cite{solit}. The resulting pattern is characterized by a dark core and a helical wave front, $E_0 = \rho_0^{1/2}(r) e^{i \psi_0}$, where $\rho_0(0)=0$ and $\psi_0 = j \theta$ with the integer $j$ being the topological charge of the vortex. The azimuthal fluid flow is thus $v_{\theta}=(1/r) \partial_{\theta} \psi_0 = c m/(k n_0 r)$. In order to create an (apparent) horizon an additional radial phase dependence is required to provide a non-zero inward radial velocity $v_r$. In any region where $v_r > c_s$, a sound wave will be swept inward by the flow and be trapped inside the horizon, that is formed where $v_r = c_s$. 

Recently, experimental evidence of an ergosphere and horizon in a photon-fluid with thermo-optical nonlinearities has been provided using a background beam with phase $\psi_0 = j \theta + 2 \pi\sqrt{r/\rz}$, where $\rz$ is an experimental parameter controlling the radial phase dependence of the beam \cite{vocke2018}. Outside the vortex core, the profile $\rho_0(r)$ asymptotes to a constant density and the flow is well approximated by the above $v_{\theta}$ and by $v_{r}=\partial_r \psi_0=-c \pi/(k n_0\sqrt{\rz r})$. Accordingly, the density fluctuations are governed by Eq. (\ref{eq11}) and the acoustic line element in polar coordinates is (up to the conformal factor $\rho_0^2/c_s^2$)
\begin{equation}
ds^2=g_{\mu\nu} dx^\mu dx^\nu \sim -c_{s}^2 dt^2 + (dr - v_r dt)^2 + (r d\theta - v_{\theta} dt)^2 \; .
\nonumber
\end{equation}

\begin{figure*}
\begin{center}
\includegraphics*[width=2.0\columnwidth]{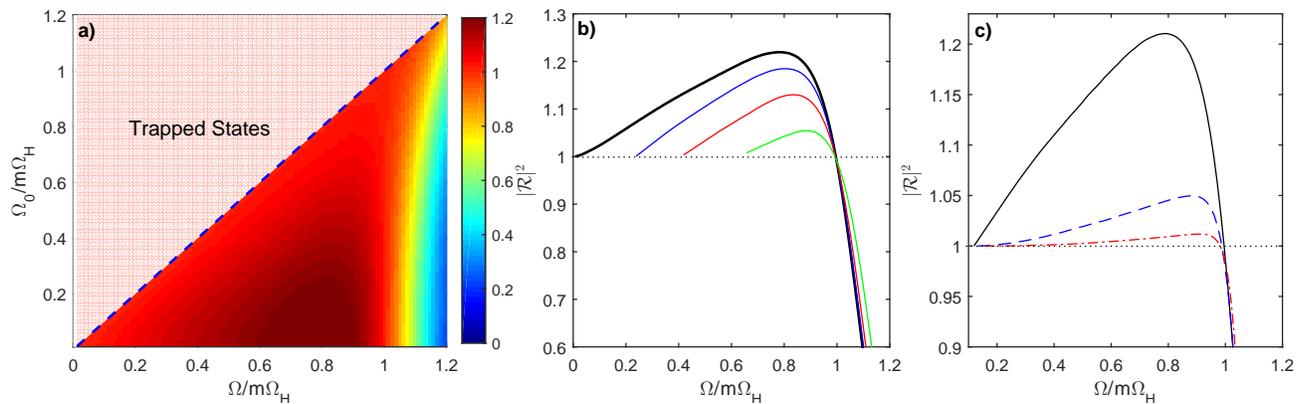}
\end{center} 
\caption{a) Reflection coefficients $|{\cal R}(\Omega)|^2$ in the ($\Omz$,$\Omega$) parameter space for $m = 1$ and b) $|{\cal R}(\Omega)|^2$ for selected values of the rest frequency: $\Omz=0$ (black), for $\Omz=0.24 \,\Omh$ (blue), for $\Omz=0.42 \,\Omh$ (red), for $\Omz=0.66 \,\Omh$ (green). c) $|{\cal R}(\Omega)|^2$ for $\Omz=0.12 \,\Omh$ and azimuthal numbers $m = 1$ (black) $m = 2$ (dashed,blue), $m = 3$ (dash-dotted,red). In all simulations $\rh=1$ and $\Omh=1.11$.}
\label{figure1}
\end{figure*}

The similarity with the equatorial slice of the Kerr geometry becomes clearer through the transformations of the time and the azimuthal coordinates 
\begin{equation}
dt \rightarrow dt + \frac{\vert v_r \vert}{(c_{s}^2 - v_r^{2})}dr ; \,\, \\ d \theta \rightarrow d\theta + \frac{\vert v_r \vert v_{\theta}}{r (c_{s}^2 - v_r^{2})}dr \; \nonumber
\end{equation}
defined in the exterior region $(\rh < r < \infty)$, where $\rh=\xi^2 /\rz$ is the location where $v_r = c_s$, i.e. the event horizon. After a rescaling of the time coordinate by $c_s$ and defining $\Omh= \frac{j \xi}{\pi \rh^2}$ the metric takes the form
\begin{eqnarray}
ds^2 \sim -\left(1 - \frac{\rh}{r} - \frac{\rh^4 \Omh^2}{r^2}\right)dt^2 + \left(1 - \frac{\rh}{r}\right)^{-1}dr^2  \nonumber\\ 
-2 \rh^2 \Omh d\theta dt + (r d\theta)^2 
\label{metric}
\end{eqnarray}
Similarly to the Kerr metric in Boyer–-Lindquist coordinates, (\ref{metric}) has a coordinate singularity at the event horizon $r=\rh$ where the radial component $g_{rr}$ goes to infinity. On the other hand, $g_{rr}$ does not contain any dependence on the azimuthal flow and therefore there is no inner horizon. The radius of the ergosphere is given by the vanishing of the temporal component $g_{tt}$, i.e. $r_\mathrm{E}=\frac{\rh}{2}(1 + \sqrt{1 + 4 \rh^2 \Omh^2})$, where $\Omh$ is the rescaled angular velocity at the horizon. The mixed term $g_{t \theta}=2 \rh^2 \Omh$ is responsible of the frame dragging due to the rotating spacetime and disappears when $\Omh=0$ (no rotation). In this case $r_\mathrm{E}=\rh$ and the metric becomes similar to Schwarzschild's. 

We now consider solutions to the Klein-Gordon equation (\ref{eq11}) with metric (\ref{metric}). We decompose the field as $\rho_1(t,r,\theta) = r^{-1/2}G(r)e^{-i\Omega t + i m \theta}$, where the integer $m$ is the winding number, $\Omega$ is normalized to $c_s$ and we introduce the dimensionless radial coordinate $\hat{r}= r/\rh$. We obtain the radial Teukolsky equation on the vortex metric
\begin{equation}
\Delta\frac{d}{d\hat{r}}\left(\Delta\frac{d}{d\hat{r}}\right) G(\hat{r}) + U(\hat{r},\hat{\Omega}) G(\hat{r}) = 0 \; ,
\label{eq12}
\end{equation}
where $\Delta=(1 - 1/\hat{r})$ and
\begin{equation}
U(\hat{r},\hat{\Omega}) = \left(\hat{\Omega} - \frac{m\hat{\Omega}_\textrm{H}}{\hat{r}^2}\right)^2 + \Delta \left(\frac{1}{4 \hat{r}^{2}}\Delta -\frac{1}{2\hat{r}^3} - \frac{m^2}{\hat{r}^2} -\hat{\Omega}_\mathrm{0}^2 \right) \; .
\label{eq13}
\end{equation}
All angular frequencies in (\ref{eq13}) have been rescaled by $\rh$ and are thus dimensionless (the frequencies 
$\Omega$ and $\Omega_\textrm{H,0}$ are indeed spatial wavenumbers due to the rescaling of the time coordinate by $c_s$ in (\ref{metric}).

In terms of the tortoise coordinate $\hat{r}^{*}$ defined as $d\hat{r}^{*} = \Delta^{-1} d\hat{r}$, Eq. (\ref{eq13}) reads
\begin{equation}
\frac{d^{2}G(\hat{r}^{*})}{d \hat{r}^*}+ U(\hat{r}^*,\hat{\Omega}) G(\hat{r}^*)  = 0 \; .
\label{eq14}
\end{equation} 
The dimensionless radial equation (\ref{eq14}) is the starting point of the analysis of the next sections.

In Kerr spacetime, superradiance and related phenomena depend on the BH angular momentum and, more importantly, on the gravitational coupling given by dimensionless product (in natural units, $G$=$c$=$\hbar$=1) of the black hole mass $M$ and the field mass $\mu$, $M\mu \equiv G M \mu/\hbar c$ \cite{dolan1}. The latter corresponds to the ratio of the event horizon size to the reduced Compton wavelength of the field, that in our potential (\ref{eq13}) is given by the dimensionless parameter $\hat{\Omega}_\mathrm{0}= r_H m_\textrm{ph} c_s/\hbar$. In the following, we characterize superradiant phenomena, quasi-bound states and scalar clouds in our system as $\hat{\Omega}_\mathrm{0}$ is varied. Unless necessary, from now on we shall omit hats in all quantities.

\section{Superradiant Scattering}

The occurrence of superradiance is demonstrated by calculating the Wronskian $\mathcal{W}$ of a solution of Eq. (\ref{eq14}) and of its complex conjugate near the horizon and at spatial infinity. 
The tortoise coordinate $r^*$ maps $\rh$ to $r^*\rightarrow -\infty$, and $r \rightarrow \infty$ to $r^* \rightarrow +\infty$. In these limits, Eq. (\ref{eq14}) translates into the following harmonic oscillator equations
\begin{eqnarray}
\frac{d^{2}G(r^{*})}{d{r^{*}}^{2}}+ \left(\Omega - m \Omh\right)^2 G(r^{*})~,~ r^*\rightarrow -\infty \label{s1}\; \\
\frac{d^{2}G(r^{*})}{d{r^{*}}^{2}}+ (\Omega^2 - \Omz^2) G(r^{*})~,~ r^*\rightarrow +\infty \label{s2} 
\end{eqnarray}
which can be easily solved to get the asymptotic solutions
\begin{eqnarray}
G(r^*) &=& {\cal T} e^{-i(\Omega - m \Omh) r^*}~,~ r^*\rightarrow -\infty \label{s3}\; \\
G(r^*) &=& e^{-i \sqrt{\Omega ^2 -\Omz^2} r^*}+{\cal R} e^{i \sqrt{\Omega ^2 -\Omz^2} r^*}, r^*\rightarrow \infty \label{s4}
\end{eqnarray}
The solution (\ref{s3}) represents a purely ingoing wave at the horizon, with is transmission coefficient ${\cal T}$, while the first and second term in (\ref{s4}) correspond respectively to an ingoing and a reflected wave with reflection coefficient ${\cal R}$. Eqs. (\ref{s3}-\ref{s4}) are limiting approximations of the solution of the full Schr\"{o}dinger equation (\ref{eq14}), the Wronskian of which is a constant independent of $r^*$. Moreover, since the potential $V(r,\Omega)$ is real (we are now considering only real frequencies $\Omega$), the complex conjugate of any solution is also a
solution. We can thus equate the Wronskian at both asymptotics $\mathcal{W}(G,G^*)\vert_{r \rightarrow \rh}=\mathcal{W}(G,G^*)\vert_{r=\rightarrow \infty}$ to find the conservation relation
\begin{equation}
1-|{\cal R}|^2=\left({{\Omega - m \Omh} \over \sqrt{\Omega ^2 -\Omz^2}}~\right)~|{\cal T}|^2~. \label{sr}
\end{equation}
Eq. (\ref{sr}) dictates that modes with frequencies in the range $\Omz < \Omega < m \Omh$, are scattered with a reflection coefficient ${\cal R} > 1$ i.e. we have superradiant amplification. Since we are working with positive frequency waves, superradiance will occur only for positive $m$, i.e., for waves that are corotating with the horizon.
Due to the presence of $\Omz$, the frequency range of superradiant scattering is reduced with respect to massless case: waves of frequency $\Omega < \Omz$ cannot propagate as they would be either divergent or exponentially suppressed at spatial infinity. The latter are the quasi-bound states that we will investigate in the next section.

The reflection coefficients $|{\cal R}(\Omega)|^2$ as a function of $\Omega$ can be explicitly calculated by numerical integration of Eq. (\ref{eq14}) with asymptotic solutions $(\ref{s1}-\ref{s2})$ and comparing the Fourier components of the incident and reflected waves. Results for different values of $\Omz$ and fixed $\Omh$ are displayed in Fig. \ref{figure1}(a,b). In the superradiant regime, $\Omz < \Omega < m \Omh$, the reflection coefficient $|{\cal R}|^2>1$ and attains a maximum just before reaching of the critical frequency $m\Omh$, beyond which it decays exponentially. The amplification is less pronounced for massive fields at all frequencies with respect to the massless case (see Fig. \ref{figure1}b): the larger is $\Omz$, the smaller is $|{\cal R}(\Omega)|^2$. In Fig. \ref{figure1}(c) we report the reflection coefficient for different values of $m$ showing that as the winding number is increased the superresonant effect is continuously decreasing.
The above phenomenology is similar to what observed in the case of Kerr spacetime \cite{reviewSR,teuko,anderson}. 

\section{Quasi-bound States}

\begin{figure}
\begin{center}
\includegraphics*[width=1.0\columnwidth]{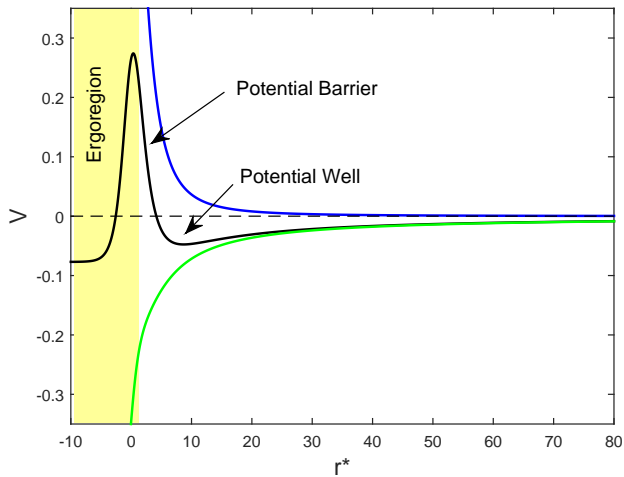}
\end{center} 
\caption{Real part of $V(r,\Omega)$ as a function of the tortoise coordinate $r^*$ for $\Omh=1.0224$ and $\Omz=0.81792$ (black), for a massless field $\Omz=0$ (blue) and $\Omh=1.0224$, and with no rotation $\Omh=0$, $\Omz=0.81792$. The shaded area indicates the ergoregion. In all plots $\Omega$ is fixed to the complex eigenfrequency $\Omega = 0.80071+2.298\times 10^{-5}i$ corresponding to a quasi-bound state (see Fig. \ref{figure3})}
\label{figure2}
\end{figure}

The instability of the Kerr spacetime to massive scalar perturbations is a consequence of a binding potential, keeping the
superradiant modes confined in the vicinity of the BH.

We then start our analysis from the potential term $U(r^*,\Omega)$ which determines the behaviour of the radial function $G$. 
We notice that defining $V(r^*,\Omega)=\Omega^2 - \Omz^2 - U(r^*,\Omega)$, Eq. (\ref{eq14}) takes the form of the 1D time-independent Schr\"{o}dinger equation
\begin{equation}
-\frac{d^{2}G(r^{*})}{d{r^{*}}^{2}}+ V(r^*,\Omega) G(r^*)  = (\Omega^\mathrm{2} - \Omz^\mathrm{2})G(r^*) \; 
\label{eq15}
\end{equation}
for a particle of "energy" ($\Omega^2 - \Omz^2$) moving in a potential $V(r,\Omega)$. Unlike the standard Schr\"{o}dinger eigenvalue problem, the potential depends on the (generally complex) eigenfrequency $\Omega$ of the modes. Nevertheless, the physical mechanism behind the superradiant amplification of trapped modes can be still interpreted in terms of the properties of $V(r,\Omega)$.
A necessary condition for the existence of the quasi-bound states is the presence of a local potential minimum. 
An example of the potential $V$ is displayed in Fig. \ref{figure1}(a), where we used the complex eigenfrequency of one of the quasi-bound states shown in Fig. \ref{figure2}(b) (black curve). Since $\Omegai \ll \Omegar$, the above picture is barely affected by $\Omegai$.
Outside the ergoregion, the curve has the typical shape of a bond interatomic potential \cite{kittel}, repulsive at short distances and attractive otherwise, and consists of a finite centrifugal barrier located in the vicinity of $r_\textrm{E}$ and of a neighboring potential well. The former is mainly related to rotation while the latter cannot form in the massless limit (see dashed curves con $\Omh=0$ e $\Omz=0$ in Fig. \ref{figure1}(a)). Trapped states in the potential well can tunnel into the ergoregion and partially enter the horizon. For $\Omegar < m \Omh$, the transmitted wave carries negative energy into the black hole and thus the reflected wave will be reflected with a larger amplitude, while remaining localized in the potential well. This leads to multiple superradiant scattering in the vicinity of the BH causing the dynamical instability.
\begin{figure*}
\begin{center}
\includegraphics*[width=2.0\columnwidth]{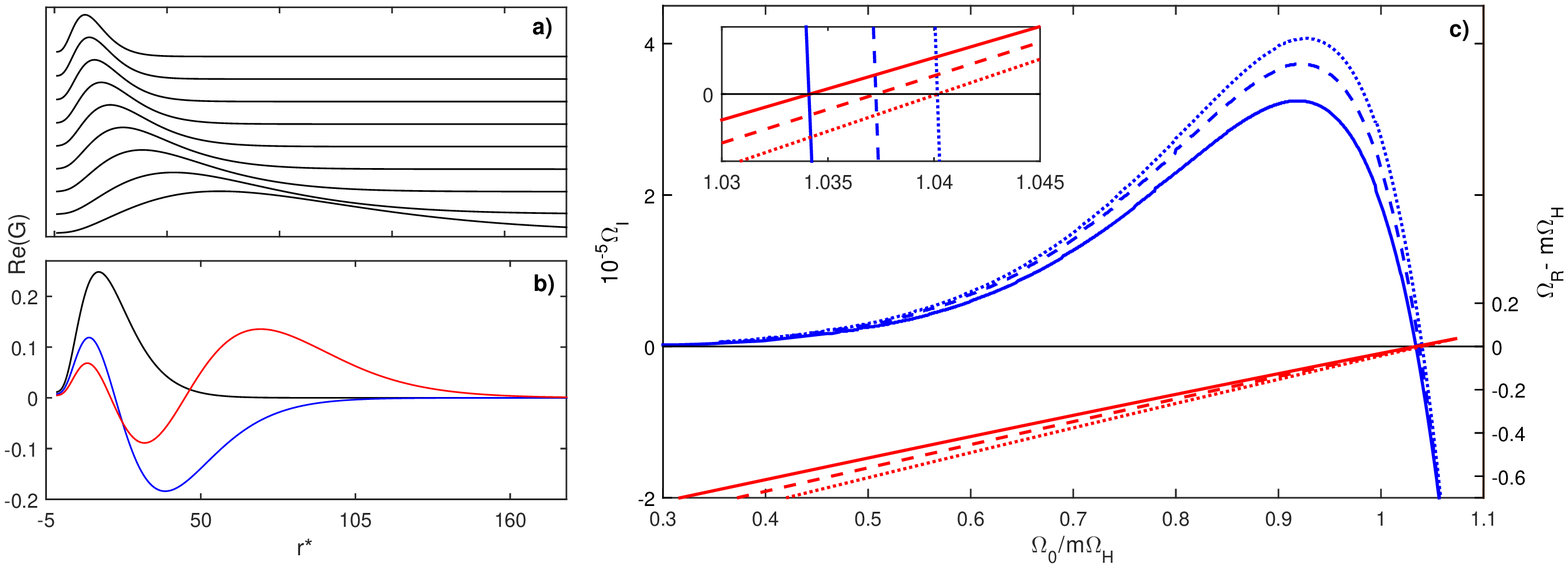}
\end{center} 
\caption{a) Radial profiles of fundamental quasi-bound states ($n=0$,$m=1$) for different values of $\Omz$ and $\Omh=1.0224$. The curves have been normalized to their maxima and an artificial offset has been added for the benefit of visualization. The values of $\Omz$ and the corresponding eigenfrequencies are reported in Table \ref{table}. b) Radial profiles of fundamental $n=0$ (black) and excited states $n=1$ (blue) and $n=2$ (red) for $\Omz=0.81792$ and $\Omh=1.0224$. Their complex eigenfrequencies are $\Omega = 0.80071525475+2.298\times 10^{-5}i$, $\Omega = 0.81001017496+9.53\times 10^{-6}i$ and $\Omega = 0.8134249684+0.00021208 i$, respectively. c) Bound state spectrum of fundamental modes ($n=0$,$m=1$) as a function of $\Omz$ and  angular velocities $\Omega_H=1.2$ (solid) $\Omega_H=1.1112$ (dashed) $\Omega_H=1.0224$ (dotted).}
\label{figure3}
\end{figure*}

\begin{table}[ht]\caption{Complex eigenfrequencies of quasi-bound states displayed in Fig. \ref{figure3}b (from down to up)}
\centering 
\begin{tabular}{c c c}\hline\hline                        
$\Omz$ \, & \, $\Omegar$ \, & \, $\Omegai$ ($\times 10^{-5}$) \\ [0.5ex] \hline                 
0.35202527 \, & \, 0.35034321 \,&  \,0.078 \\
0.43203101 \, & \,  0.42909303 \, & \,  0.080 \\
0.52803790 \, & \,  0.52298580 \, & \,  0.230 \\
0.62582270 \, & \,  0.61784984 \, & \,  0.500 \\
0.73338598 \, & \,  0.72119458 \, & \,  1.080 \\
0.83917135 \, & \,  0.82167984 \, & \,  2.040 \\
0.93517824 \, & \,  0.91176974 \, & \,  3.090 \\
1.0498531  \, & \, 1.0178151 \, & \, 3.640 \\
1.1583054  \, & \, 1.1163115 \, & \,  -0.462  \\ [1ex]       
\hline
\end{tabular}\label{table}
\end{table}

The above states are described by quasi-bound (non-stationary) modes behaving as purely ingoing waves at the horizon (as measured by a comoving observer) and exponentially-decaying (bounded) solutions at spatial infinity. In terms of the tortoise coordinate, their asymptotic radial profiles are written as
\begin{eqnarray}
G(r^*) &\sim & e^{-i(\Omega - m \Omh) r^*}~,~ r^*\rightarrow -\infty \label{s5}\; \\
G(r^*) &\sim & e^{-\sqrt{\Omz^2 -\Omega^2} r^*}~,~ r^*\rightarrow +\infty \label{s6}
\end{eqnarray}
For a given $\Omz$, the radial equation (\ref{eq15}) [or, equivalently Eq. (\ref{eq14})] together with the above boundary conditions gives rise to a discrete spectrum of complex eigenfrequencies $\{\Omega(\Omz)\}=\Omegar + i \Omegai$. The imaginary part $\Omegai$ sets the growth ($\Omegai > 0$) or decay ($\Omegai < 0$) rate of the amplitude of the scalar field.  
We remind that in addition to bound-states, the problem admits also a discrete set of resonances corresponding to outgoing waves at spatial infinity $ \sim e^{\sqrt{\Omz^2 -\Omega^2} r^*}$, which are the acoustic analogue of the so-called quasi-normal modes \cite{simone,kono,hod5}. In the following we focus on quasi- and purely-bound state solutions.

Fixed the black hole angular frequency $\Omh$ and the rest frequency of the massive field $\Omh$, quasi-bound states are uniquely identified by two “quantum” numbers: the winding number of the field $m$ and the non-negative $n$ corresponding to the node number of $G(r^*)$. Notice that although the topological charge of vortex $j$ is also quantized, it does not uniquely determine the state, as $\Omh$ can be continuosly varied e.g. by changing the location of the horizon.

To calculate the quasi-bound state spectrum we numerically solve Eq. (\ref{eq14}) implementing a Numerov integration scheme and one-parameter shooting method based on bisection procedure to find the eigenvalues. In our calculations we keep $\Omh$ fixed and treat $\Omega_R$ as the eigenvalue to be determined, whereas the parameters $\Omegai$ and $\Omz$ are continuously scanned. First we perform forward integration, starting from $r = r_H+\delta r$ until a suitable chosen matching point $r_m$, and then a backward one, starting from $r = r_{max}$ until that point. The trial value of $\Omega _R$ is properly adjusted in order to get the pre-determined number $n$ of nodes and the matching between the forward and backward solutions. The latter is evaluated by comparing their logarithmic derivative at $r_m$.  

The radial profiles of quasi-bound modes with $n=0$ and $m=1$ for different values of $\Omz$ are illustrated in Fig. \ref{figure3}(a). The solutions are strongly reminiscent of the ground-state radial wavefunctions in hydrogenic potentials. We observe that the position of their maxima increases with $\Omz$ and, contextually, the correspondent profiles are less localized (i.e. they extend over a larger portion of space). This behaviour is consistent in the framework of a quantum mechanical interpretation of $G$ as an atomic orbital, in which the mode "position" related to the Bohr radius of the massive particle is proportional to its Compton wavelength, thus scaling as $1/\Omega_\mathrm{0}$.

In Fig. \ref{figure3}(c) we report the spectrum of the fundamental quasi-bound state for ($n$,$m$)=($0$,$1$). The plot clearly shows that a dynamical instability ($\Omegai > 0$) occurs within the superradiant regime, $\Omegar < \Omz < m \Omh$.
Similarly to the curves of the reflection coefficient $\mathcal{R}$ (see Fig. \ref{figure1}), the growth rate displays a clear maximum close the critical frequency ($\Omegar \lesssim \Omz \sim m\Omh$) and increases with the BH angular velocity. Instead, modes with $\Omegar > m \Omh$ are decaying ($\Omegai < 0$), meaning that the field is infalling into the horizon. When $\Omegar= m \Omh$ the growth rate is equal to $1$. In these conditions the trapped modes are neither growing nor vanishing. This is precisely the equilibrium condition for the existence of stationary clouds that we will discuss in the next section. 
At lower values of $\Omz$ the growth rate is continuosly decreasing, approaching the value $\Omegai=0$ as $\Omegar \rightarrow \Omz$. For $\Omega > \Omz$ quasi-bound states no longer exist and the system enters the regime of supperradiant scattering (see Sec. V). The above spectrum is fully consistent to what observed for massive scalar fields on Kerr spacetime \cite{dolan1,huang2,dolan2}.

In Fig. \ref{figure3}(b) we show quasi-bound states for $m=1$ and $n=0,1,2$. We find that increasing the number of nodes 
the profiles are less localized and their center-of-mass shifts to higher values of $r^*$. These solutions have been found for three complex eigenfrequencies, the real part of which increases with $n$ (see caption of Fig. \ref{figure3}). Therefore, when all the other parameters are fixed, modes with a higher number of nodes are more energetic with respect to the fundamental and can be actually interpreted as excited states of the field. 


\begin{figure*}
\begin{center}
\includegraphics*[width=2.0\columnwidth]{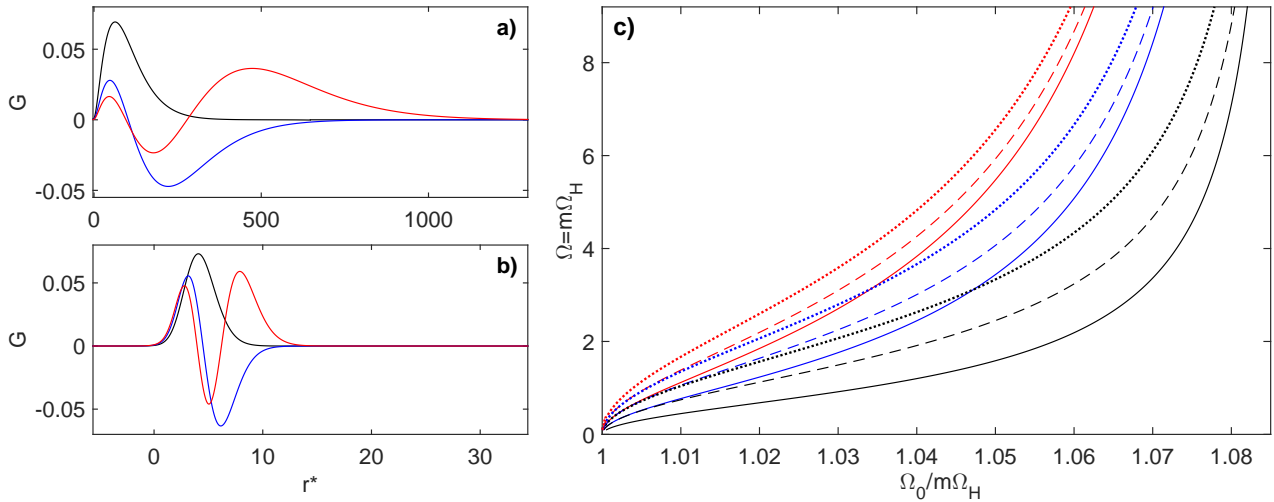}
\end{center} 
\caption{Radial profiles of fundamental $n=0$ (black) and excited scalar clouds $n=1$ (blue) and $n=2$ (red) for $m=1$ and a) $\Omz=0.2791$ and b) $\Omz=7.2441$. The corresponding eigenfrequencies are a) $\Omega =0.277952093$ ($n=0$), $\Omega = 0.278675218$ ($n=1$), $\Omega = 0.278881324$ ($n=2$) and b) $\Omega = 6.71066794$ ($n=0$), $\Omega = 6.79451843$ ($n=1$) and $\Omega = 6.86120254$ $n=2$.  Existence lines in the $(\Omh, \Omz)$ parameter space for clouds: ($n$,$m$)=($0$,$1$) (solid-black);($n$,$m$)=($0$,$2$) (dashed-black);($n$,$m$)=($0$,$3$) (solid-black);($n$,$m$)=($1$,$1$) (solid-blue);($n$,$m$)=($1$,$2$) (dashed-blue);($n$,$m$)=($1$,$3$) (dotted-blue);($n$,$m$)=($2$,$1$) (solid-red);($n$,$m$)=($2$,$2$) (dashed-red);($n$,$m$)=($1$,$3$) (dotted-red).}
\label{figure4}
\end{figure*}

\section{Scalar Clouds}

As discussed in the previous section, at the superradiant instability threshold the phonon field is in equilibrium with the BH forming a pure bound state, i.e. $\Omegai=0$. We thus solve numerically Eq. (\ref{eq14}) fixing $\Omegar = \Omega = m \Omh$. 
In this case, the radial function admits a power series expansion close to the horizon 
\begin{equation}
G \sim 1 + c_1(r-1) + c_2(r-1)^2 +...,
\label{ex}
\end{equation}
where the coefficients can be found replacing (\ref{ex}) into Eq. (\ref {eq14}) and solving it order by order in terms of powers of (r-1). We adopt a numerical method similar to the one used to calculate quasi-bound states: starting with the near horizon expansion (\ref{ex}), we fix all quantum numbers and scan $\Omz$ to find the eigenfrequencies $\Omega$ for which $G$ remains bounded, as prescribed by the condition (\ref{s6}). In our scheme we have truncated the expansion at the second-order. Solutions with the right asymptotic behavior exist only for a discrete set of $\Omega$ corresponding to number of nodes $n$ of the radial function $G$.

In Fig. \ref{figure4}(a-b) we plot the radial dependence of fundamental and excited clouds for two different rest frequencies. The profile is always bounded at the horizon and decrease exponentially as $r^* \rightarrow \infty$. 
Similarly to quasi-bound states, scalar clouds with larger $\Omega_0$ exhibit a smaller orbital radius and a smaller width of the associated wave profile.


The above simulations allow to construct the existence line of clouds in the ($\Omh, \Omz$) space.
Such lines for the fundamental $n = 0$ and excited modes $n = 1, 2$ and different azimuthal numbers $m$ are displayed in Fig. \ref{figure4}(c). Since these lines demarcate the threshold of the superradiant instability, the area below (above) a given line contains unstable (stable) solutions against the corresponding mode.

We observe that the existence lines shift towards higher values of $\Omh$ for higher $n$ and $m$, which is in agreement with the interpretation that excited configurations require a larger background rotation for equilibrium.
All the lines are converging as $\Omega \rightarrow \Omz$. However, this is just an asymptotic limit since in that point
$\Omz=m \Omh$ and accordingly to conditions (\ref{s5}-\ref{s6}) the radial function $G$ would become a fully delocalized (plane) wave.

We now comment on how close to the horizon the clouds can be localized. We first notice that for any set of parameters the radial profile of the fundamental mode $n = 0$ attains its maximum in correspondence of the minimum of the potential well.
A rough estimate of its position can be thus obtained by finding the local minimum of $V(r,\Omega)$.

In order to obtain more elucidating expressions, we approximate the potential considering only the first leading terms in 1/r, i.e.
\begin{equation}
V(r,\Omega)=-\frac{\Omz^2}{r} + \frac{w}{r^2}
\label{eq16}
\end{equation}
where we used $\Omega=m\Omh$ and we defined $w=m^2+2\Omega^2 - 1/4$. Eq. (\ref{eq16}) closely resemble the Kratzer molecular potential with an attractive Coulombian and a repulsive inverse-square term \cite{kratzer}. The minimum is found at $r_\textrm{cl}=2w/\Omz^2$. The dependence of $r_\textrm{cl}$ from BH and field parameters can be inferred with the help of the existence lines in Fig. \ref{figure4}(c). As $\Omz \rightarrow m \Omh$, $\Omega \rightarrow 0$ and 
\begin{equation}
r_\textrm{cl}=\frac{2}{\Omh^2} - \frac{1}{2 m^2 \Omh^2} \geq \frac{3}{2 \Omh^2} \equiv r_\textrm{min}
\label{eq17}
\end{equation}
We can see that $r_\textrm{cl}$ increases as the angular frequency descreases, diverging as $\Omh \rightarrow 0$. This behaviour is in agreement to what observed in the case of Kerr BHs and is consistent with the fact that clouds are supported only in the presence of rotation. Similar considerations apply to the limit case $\Omz \rightarrow 0$, keeping $\Omh$ fixed: without any other confinement mechanism, clouds can form only in the presence of massive scalar fields.

\begin{figure}
\begin{center}
\includegraphics*[width=1.0\columnwidth]{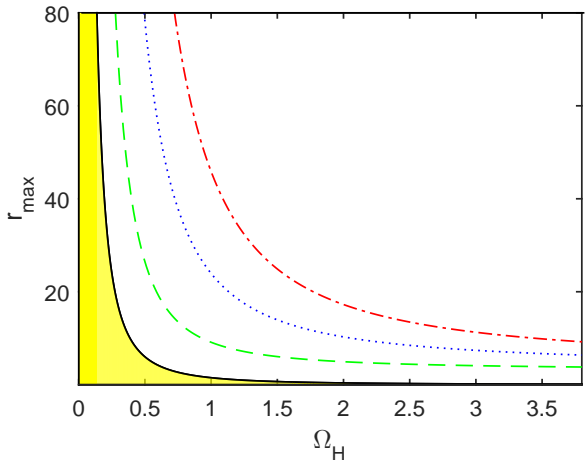}
\end{center} 
\caption{Effective positions of the fundamental cloud solution $(n = 0, m=1)$ (dashed-green) and the first two excited states $(n = 1, m=1)$ (dotted-blue) and $(n = 2, m=1)$ (dashed-dotted-red) as a function of $\Omh$. The solid-black curve is the function $r_\textrm{min}(\Omh$) as defined in (\ref{eq17}).}
\label{figure5}
\end{figure}

In Fig. \ref{figure5} we plot the effective position of cloud solutions, calculated as the radial coordinate for which the function $2\pi r |G(r)|^2$ attains its maximum value \cite{hod-hair}, as a function of $\Omh$. As expected, the excited states $n = 1, 2$ are located at larger distances from the horizon with respect to the fundamental mode $n = 0$ (solutions with higher azimuthal numbers $m$ are even farther). All curves are in agreement with the inequality (\ref{eq17}) that thus provide a lower bound for the clouds position.

Unlike the Kerr metric, acoustic BHs have no mathematical upper limits on the angular velocity \cite{berti,marino2009}. On the other hand, a soft bound is placed by $K \ll K_c$, i.e. the long-wavelength regime in which the massive Klein-Gordon equation well describe the phonon dynamics. This condition implies an upper bound for the (dimensionless) angular frequency $\Omh = \rh K_c /m$ that would correspond to an absolute lower limit for the position of cloud solutions.
Incidentally, for $\Omh \sim 1$,  Eq. (\ref{eq17}) is in perfect agreement with the “no-short hair” conjecture \cite{noshort}, stating that for static and spherically symmetric BHs the region of non-asymptotic behavior of the BH hair (hairosphere), of which the scalar clouds are the linear seeds, must extend beyond $3/2$ the horizon radius \cite{hod-hair}.

We remark that these simple considerations are by no means an attempt of showing an acoustic analogue of no-short hair conjecture, nor even limited to the (linearized) cloud solutions. A stronger analogy should be established in terms of the position of circular null geodesics of the vortex metric, that should provide a more general measurement of the minimal hair extension \cite{hod6}. Nevertheless, a lower bound for clouds in our system indeed exists and is simply related to the position of the potential well. As a second comment, we observe that the limit frequency $\Omh = \rh K_c /m$ only separates two different regimes, namely the Lorentz-invariant phonon regime and the high-energy regime of Bogoliubov excitations. Recent works in local photon-fluids have demonstrated the occurrence of a resonant amplification phenomenon for Bogoliubov excitations, which reduces to the standard phonon superradiance in the long-wavelength limit \cite{prainSR}. Remarkably, the superradiant regime $\Omega < m \Omh$ remains unaffected independently of the energy of the excitation, and the effects of the breakdown of the long-wavelegth approximation manifest only in a reduced amplification factor. While the generalization of these results to the case of our photon-fluid is not straightforward, we would not expect strong qualitative differences, in particular for scalar clouds solutions, which exist right at the instability thereshold $\Omega = m \Omh$. All this will be the subject of a forthcoming investigation.

\section{Conclusions and Perspectives}

Quantum fluids of light such as exciton-polaritons BECs and more recently photon-fluids have offered alternative platforms for analogue gravity investigations. Recent experiments in these systems provided evidence of acoustic horizons in 1D flows \cite{elazar,nguyen} and rotating BH geometries \cite{vocke2018,jacquet} in (2+1)-dimensions.

Here, we have theoretically investigated a photon-fluid with both local and nonlocal thermo-optical interactions, the fluctuations of which behave as a massive Klein-Gordon field. In the presence of suitable vortex flows, superradiant instabilities are shown to arise due to the existence of phonon states which are prevented from radiating their energy to infinity. At the superradiant instability thereshold, these modes give rise to purely bound states in synchronous rotation with the horizon, which exactly represent the acoustic counterpart of stationary scalar clouds around Kerr BHs.
This generalizes previous studies to the case of massive phonons and pave the way for observing superradiant instabilities and massive scalar clouds configurations in the laboratory. Since most analogue gravity models are dealing with massless excitations, this is one of the very few systems in which these phenomena could be investigated and, possibly in the near future, experimentally observed. 

Interestingly, scalar clouds have been related to the existence of Kerr BHs with scalar hair \cite{herdeiro1}: these solutions of the linear Klein-Gordon equation are actually the seeds of a new family of solutions of the fully nonlinear Einstein-Klein-Gordon system, which correspond to Kerr-like geometries deformed by the cloud backreaction \cite{herdeirobr}.
As also remarked in \cite{benone-ac}, the specific form of the equations ruling the background is thus crucial to relate linear clouds around Kerr BHs to (nonlinear) scalar hairy BHs. 
As in all analog models, our backreaction is encoded in a set of nonlinear equations [cf. Eq. (\ref{eq3}-\ref{eq4})] which have neither the form nor the symmetries of Einstein’s equations. 
Nevertheless, an interesting feature of our system is that when the first nonlinear corrections are taken into account, the phonon-modes behave as a \emph{self-gravitating} quantum system with an effective Schr\"{o}dinger-Newton dynamics. Similarly to the Newtonian limit of General Relativity (GR), the potential in the Poisson equation is related to the background geometry experienced by the particles propagating on it (namely to the 00-component of the metric), and the source term depends of the phonon mass-density distribution. Although in ($2+1$)-dimensions the weak-field limit of GR does not correspond to 2D Newtonian gravity (thus the above dynamics could not be considered either as a limit of Einstein's gravity), it would be anyway fascinating to explore the weakly nonlinear regime, where the backreacting clouds could give rise to scalar field configurations frequency-locked to the vortex background, potentially interesting for future analogue gravity investigations.

\end{document}